\documentclass[runningheads]{llncs}
\usepackage{xcolor}
\usepackage{soul}
\usepackage{hyperref}
\usepackage[super]{nth}

\usepackage{booktabs}

\usepackage{amsmath}

\DeclareMathOperator*{\argmin}{arg\,min}

\usepackage{graphicx}
\hypersetup{colorlinks,allcolors=black}

\begin{document}
\title{Deformable image registration with deep network priors: a study on longitudinal PET images
\thanks{This work is partially financed through "Programme op\'era-tionnel regional FEDER-FSE Pays de la Loire 2014-2020" n\textsuperscript{o}PL0015129 (EPICURE); and by Lilly and AstraZeneca for the clinical trial support.}
}
\titlerunning{Deformable image registration with deep network priors}
%
\author{Constance Fourcade$^{1,2}$  \qquad Ludovic Ferrer$^{4,5}$\textit{ PhD} \\ 
No\'emie Moreau$^{2}$ \qquad  Gianmarco Santini$^{2}$\textit{ PhD}\\ 
Aislinn Brennan$^{2}$  \qquad Caroline Rousseau$^{3,5}$\textit{ MD PhD} \\
Marie Lacombe$^{5}$\textit{ MD} \qquad Vincent Fleury$^{5}$\textit{ MD} \\ 
Mathilde Colombi\'e $^{5}$\textit{ MD} \qquad Pascal J\'ez\'equel$^{4,5}${ PhD} \\ 
Mario Campone$^{3,5}$\textit{ MD, PhD} \qquad Mathieu Rubeaux$^{2}$\textit{ PhD}\\ 
Diana Mateus$^{1}$\textit{ PhD}}
\authorrunning{Fourcade. et al.}
%

\institute{1. Ecole Centrale de Nantes, LS2N, UMR CNRS 6004, Nantes, France \\
           2. Keosys Medical Imaging, Saint Herblain, France \\
           3. University of Nantes, CRCINA, INSERM UMR1307, CNRS-ERL6075, Nantes, France \\
           4. University of Angers, CRCINA, INSERM UMR1307, CNRS-ERL6075, Angers, France \\
           5. ICO Cancer Center, Nantes - Angers, France \\
\email{constance.fourcade@ec-nantes.fr}}
\maketitle              
\begin{abstract}

This paper proposes a novel approach for the longitudinal registration of PET imaging acquired for the monitoring of patients with metastatic breast cancer. 
Unlike with other image analysis tasks, the use of deep learning has not significantly improved the performance of image registration. With this work, we propose a new registration approach to bridge the performance gap between conventional and deep learning-based methods: Medical Image Registration method Regularized By Architecture (\texttt{MIRRBA}).
\texttt{MIRRBA} is a subject-specific deformable registration method which relies on a deep pyramidal architecture to parametrize the deformation field. Diverging from the usual deep-learning paradigms, \texttt{MIRRBA} does not require a learning database, but only a pair of images to be registered that is used to optimize the network’s parameters. 
We applied \texttt{MIRRBA} on a private dataset of 110 whole-body PET images of patients with metastatic breast cancer. We used different architecture configurations to produce the deformation field and studied the results obtained. We also compared our method to several standard registration approaches: two conventional iterative registration methods (ANTs and Elastix) and two supervised deep learning-based models (LapIRN and Voxelmorph). 
Registration accuracy at a global and local level was evaluated using the detection rate and the Dice score respectively, while the realism of the registration obtained was evaluated using Jacobian’s determinant. The ability of the different methods to shrink disappearing  lesions was also computed with the disappearing rate. 
MIRRBA significantly improved the organ and lesion Dice scores of Voxelmorph by 6\% and 52\% respectively, and of LapIRN by 5\% and 65\%. Regarding the disappearing rate, MIRRBA more than doubled the score of the best performing conventional approach ANTs.
In this paper, we also demonstrate the regularizing power of deep architectures and present new elements to understand the role of the architecture in deep learning methods used for registration. 

\keywords{Longitudinal image registration  \and PET \and Deep image prior \and Breast cancer}
\end{abstract}

\section{Introduction}

Medical image registration is the precise overlaying of a fixed and moving image. It is notably used to create patient models (inter-patient monomodal registration), exploit information from different modalities acquired for a single patient (intra-patient multimodal registration), and to monitor tumor evolution (longitudinal registration).  

However, the accuracy of image registration is a long-standing fundamental problem in medical image analysis (\cite{Maurer1993,Fu2020}). In the past 30 years, registration methods have considered registration as an optimization problem: the goal is to minimize a dissimilarity term between moving and fixed images. It is usually measured with mean square error (MSE), mutual information (MI) or normalized cross-correlation (NCC). In addition, a term enforcing smooth and realistic transformations is often used as a regularizing term. 

Most conventional methods manage registration by parameterizing the transformation between the fixed and moving image. They use for instance discrete cosine transforms (\cite{Friston1995}), 3D Fourier series (\cite{Christensen2007}), cubic B-splines (\cite{Klein2010}) or optimized velocity fields (\cite{Avants2009}). The parameters of the model are typically optimized though efficient second-order minimization (\cite{Vercauteren2007a}) or stochastic gradient descent (SGD) (\cite{Klein2009}), within a single- or multi-level optimization strategy. The minimized cost function combines dissimilarity and regularization terms.

Recent deep learning (DL) supervised registration methods (\cite{Fu2020}) also rely on dissimilarity measurements and regularization terms that can be optimized with variants of SGD.  However, there are two key differences compared to conventional approaches. DL methods require a training stage with a training database that gives the model prior knowledge regarding deformations.  They also use a different type of transformation model, with convolutional neural networks (CNN) used as an over-parameterized but structured model of the deformation field.

Despite recent developments in DL-based registration (\cite{Chen2021}), conventional methods still perform better and obtain more accurate results in most applications (\cite{Heinrich2020}). In addition, even though trained networks are faster than conventional methods, training requires large databases which often restricts their use to a specific therapeutic area.

With this work, we propose a new registration method formalized as a conventional registration approach, with a deformation field modeled by an untrained deep pyramidal network. We named our approach \texttt{MIRRBA} for \textit{Medical Image Registration method Regularized By Architecture}. 

We applied \texttt{MIRRBA} on longitudinal PET images acquired for the evaluation of treatment response in patients with metastatic breast cancer and compared the results to registrations obtained with standard approaches. We also studied the impact of different architecture configurations (14 configurations) on the deformation field. 

The contributions of this paper are i) the proposition of a novel registration method regularized by architecture MIRRBA, ii) an extensive comparative study of the effects of different network components on the deformation field, and iii) a solution to register whole body PET images both at a global and local level without the need for prior registration to facilitate the simultaneous monitoring of multiple lesions.

\section{Related work}

\paragraph{Automatic longitudinal lesion monitoring}

Automatic approaches for lesion segmentation and/or image registration have been developed over the years to help with cancer monitoring and treatment response assessments. 

\cite{Necib2011} used affine registration to align baseline and follow-up PET images, before subtracting them to identify tumor voxels showing significant changes between the two scans. However, since affine registration is a global method, it mainly performs well for single localized tumors. In a situation where there were multiple lesions to assess, \cite{Hsu2020} first segmented all lesions on liver-centered CT images, then used a longitudinal correspondence module to find matching pairs of lesions from the segmentation maps, and finally computed the lesions’ evolution. Even if promising, multi-staged methods suffer from error propagation between stages. \cite{Chassagnon2020} removed the need for a prior segmentation step and relied on a conventional registration algorithm between longitudinal CT pairs to obtain the deformation fields and their Jacobian determinants. This information was then used to train a classifier network assessing systemic sclerosis interstitial lung disease. 

Image registration can be a key step in lesion monitoring. As our long-term goal is to use registration results to develop tools for the monitoring of metastatic breast cancer, this work focuses on registration and on how to improve its accuracy. We present hereafter the most used registration approaches and the good practices which contribute to make them successful.

\paragraph{Established registration tools} 

There is a large body of literature on conventional registration methods (\cite{Klein2009a,Sotiras2013}). Some of today’s most used registration tools include the Demons method (\cite{Vercauteren2007}) using non-parametric diffeomorphic displacement fields, the Elastix toolbox (\cite{Klein2010}) based on cubic B-splines, and the Advanced Normalization Tools (ANTs) (\cite{Avants2009}) parametrizing the velocity field and relying on bi-directional diffeomorphisms. We compared our proposed approach to the last two methods, as they perform well with different datasets (\cite{Klein2009a}) and their pyramidal coarse-to-fine optimization has inspired recent works on DL-based registration which is presented hereafter.

\paragraph{DL methods}
\label{sec:DL_methods}

In recent years, different types of DL-based registration approaches have been proposed (\cite{Chen2021}). With the monitoring of metastatic breast cancer, deformation fields are not available and are difficult to obtain. We therefore focused on unsupervised registration methods. We categorize as unsupervised methods which do not require ground truth deformation fields, segmentations, or landmarks. Typically, they learn by minimizing a dissimilarity term between the fixed and the wrapped moving image. Among the unsupervised registration methods, various CNN architectures have been proposed. The widely used U-Net (\cite{Ronneberger2015}) inspired the reference Voxelmorph network first trained on brain MRIs (\cite{Balakrishnan2018}). \cite{Stergios2018} added dilated convolutions to the encoder path of Voxelmorph to capture a wider field of view from lung MRI images. Later, \cite{DeVos2019} proposed a U-Net-based cascade network applied to cardiac cine MRI and chest CT data to perform affine and deformable registrations in stages, at the cost of an increase in complexity. On 3D PET images, \cite{Li2021} also proposed an iterative DL-based registration method to reconstruct motion compensated PET images. While still using a U-Net-shaped architecture, \cite{Eppenhof2020} proposed a coarse-to-fine training mimicking the best performing conventional approaches. With a similar training strategy, \cite{Mok2020} won the Learn2Reg 2021 MICCAI challenge\footnote{ \href{https://learn2reg.grand-challenge.org/}{https://learn2reg.grand-challenge.org/}} using a pyramidal network named LapIRN. Contrary to the above cited methods, \cite{Mok2020} did not only impose a smoothness constraint on the deformation field through its gradient, but also enforced diffeomorphic transformations using stationary vector fields under the Log-Euclidean framework, as in \cite{Dalca2019}.

Despite recent efforts, when using DL-based registration techniques no major gain in registration performance was reported. Conventional iterative optimization methods still yield better accuracy in many tasks such as inter-patient alignment or intra-patient lung motion registration (\cite{Heinrich2020}). Large databases would be needed to learn the network parameters and produce accurate deformation fields. Moreover, we can argue that the generalization ability of the trained network is questionable when the deformation patterns do not repeat consistently across the dataset. 

\paragraph{Deep Image Prior registration}
To alleviate the dependency to a dataset and the need for repeated deformation patterns, we adapted the Deep Image Prior (DIP)  framework (\cite{Ulyanov2020}) to propose a learning-free method for deformable medical image registration. Contrary to standard DL-based approaches, DIP does not learn from a database but relies on a single image. It uses a deep architecture not to summarize the information across samples but as a prior. Effectively, the architecture plays the role of a parametric model in an optimization problem restricting the solution space. While DIP was initially designed for denoising and inpainting tasks (\cite{Ulyanov2020}), we adapted it here to medical image registration.

\cite{Laves2019} suggested image registration as a potential application of DIP, with preliminary results in the context of 2D brain MR images. DIP has also been used on medical data for the reconstruction of CT and PET images (\cite{Gong2019,Baguer2020}). To perform these reconstructions, an untrained deep network was used as a denoiser and iteratively optimized by a conventional algorithm. Apart from the use of an initial reconstruction and classical regularization, images from other modalities were also given as input to condition the network output.

With this work, we establish a link between conventional, deep learning and DIP-based approaches. In particular, we focus on the role of the untrained network in parameterizing the displacement field, showing that each architecture induces an implicit regularization when used within a conventional optimization scheme. Similar observations have been made in the context of DL for inverse problems \cite{Lucas2018} and \cite{Dittmer2020}. As a result, well-structured architectures (e.g. \cite{Mok2020}) provide better over-parameterizations for the deformation fields, both in the supervised and in the untrained cases. We also investigate the role of the input (random, moving or moving and fixed images) and potential interactions of the untrained network with conventional supervised approaches.

\begin{figure*}[t]
 \centering
 \includegraphics[width=\textwidth]{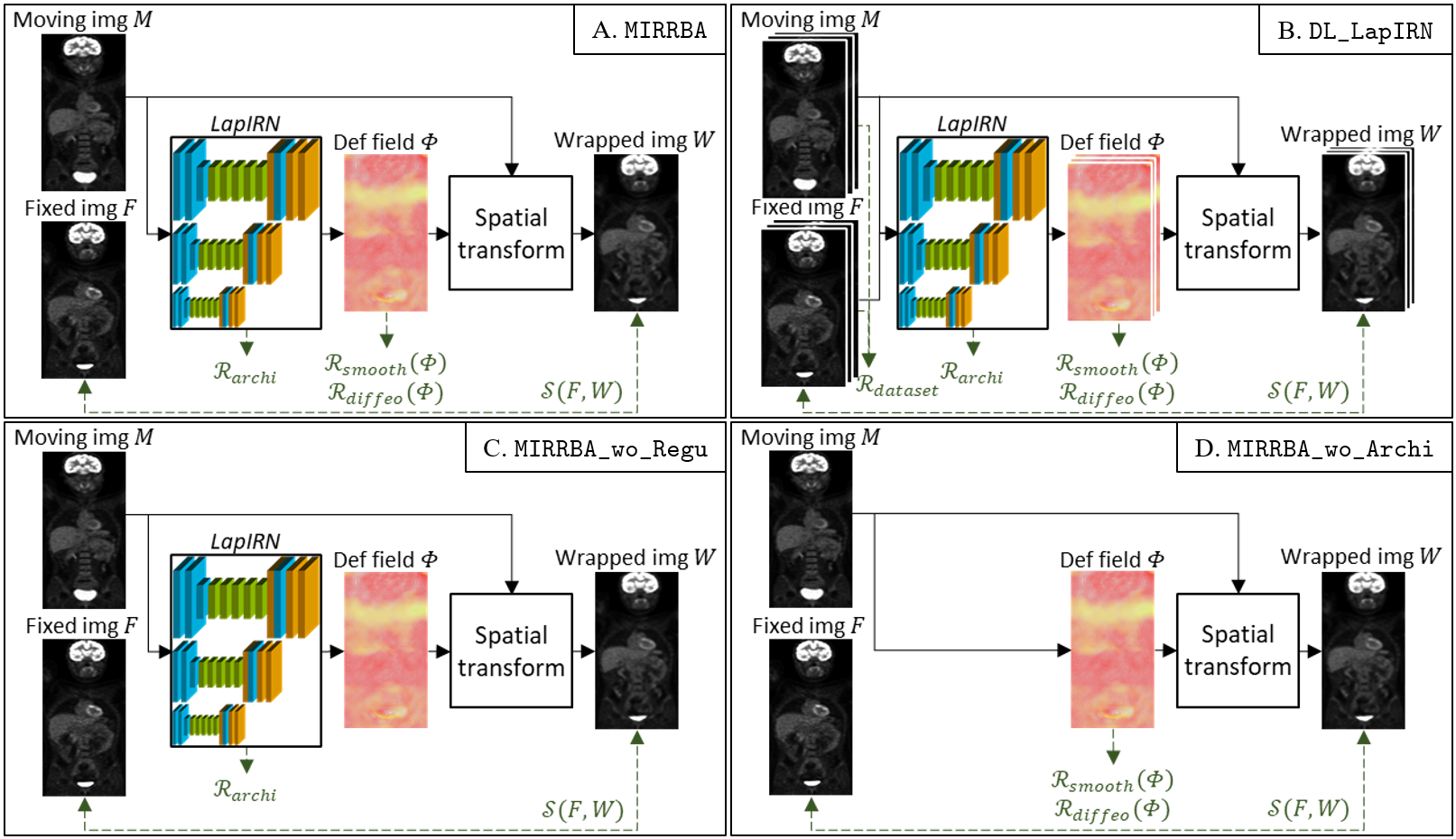}
 \caption{Overview of A. \texttt{MIRRBA}, B. \texttt{DL\_LapIRN} and C. \texttt{MIRRBA\_wo\_Regul} and D. \texttt{MIRRBA\_wo\_Archi} methods. The LapIRN architecture is visible with the encoder path (blue), the residual blocks (green) and the decoder path (orange with a blue layer from the encoder path). Best viewed in color.}
 \label{fig:methods}
\end{figure*}

\section{Method}

\subsection{Image registration}
\label{sec:img_reg}

Given a pair of fixed and moving $(F$, $M)$ images, deformable image registration aims to estimate a dense deformation field $\Phi$ such that the wrapped image $W = M(x + \Phi(x))$ is aligned with $F$ for each voxel $x$. Given a metric $\mathcal{S}(.,.)$ measuring the dissimilarity between two images, conventional registration methods choose a parametrization of transformation and optimize it through a cost function of the form:
\begin{equation}
    \label{eq:reg_1}
    \underset{\Phi}\argmin \hspace{1mm} \mathcal{S}(F, W)
\end{equation}
However, the problem stated Eq.~\ref{eq:reg_1} is ill-posed: it belongs to the nonlinear transformation class (\cite{Myronenko2009}). By explicitly constraining the transformation, parametric approaches can become well-posed (\cite{Rueckert1999}), at the cost of limiting the range of admissible transformations. 
Constraining an over-parameterized or complex deformable transformation is effectively done by adding a regularization term $\mathcal{R}_{smooth}(\Phi)$ to the objective function Eq.~\ref{eq:reg_1}. Regularization terms incorporate priors, which if correctly chosen, guide the optimization towards better estimates of the  deformation fields.
Weighted by $\lambda_{smooth}$, the most common regularization term enforces smoothness onto the displacement field by penalizing the spatial derivatives of $\Phi$.
\begin{equation}
    \label{eq:reg_2}
    \underset{\Phi}\argmin \hspace{1mm} \left\{ \mathcal{S}(F, W) + \lambda_{smooth} \mathcal{R}_{smooth}(\Phi) \right\}
\end{equation}
DL-based registration approaches (\cite{Fu2020}) adapt the cost function in Eq.~\ref{eq:reg_2} by modeling $\Phi$ to be  the output of a CNN trained on a dataset (See Eq.~\ref{eq:reg_3}).
The inputs of this network are the $(F$, $M)$ pair, and the warping operation to get $W$ is performed using a spatial transformer layer (\cite{Jaderberg2015}), handling the transformation, the sampling and the interpolation. Regarding the regularization,  DL-registration based methods continue relying on a term enforcing the smoothness of the transformation $\mathcal{R}_{smooth}$, as done in conventional approaches. Smoothness is often completed by $\mathcal{R}_{diffeo}$, which enforces the diffeomorphism of the transformation by penalizing the determinant of the Jacobian negative values. 
Regularization terms are weighted respectively by $\lambda_{smooth}$ and $\lambda_{diffeo}$. Moreover, we state that two additional regularizing priors are also implicitly added in the DL setup and influence the predicted deformation field $\Phi$: the first one $\mathcal{R}_{dataset}$ is induced by training on a domain-specific dataset, while the second one $\mathcal{R}_{archi}$ is entailed by the network architecture choice (see Fig.~\ref{fig:methods}.B). Due to the dataset dependency, one limitation of DL-based registration methods is the difficulty to generalize across organs or modalities.
\begin{equation}
    \label{eq:reg_3}
    \underset{\Phi(\mathcal{R}_{dataset}, \mathcal{R}_{archi})}\argmin \hspace{1mm} \{ \mathcal{S}(F, W) + \lambda_{smooth} \mathcal{R}_{smooth}(\Phi) + \lambda_{diffeo}\mathcal{R}_{diffeo}(\Phi) \}
\end{equation}

Next, we recall the deep image prior concept, which removes the dataset dependency in Section \ref{sec:dip}, and present our method for registration based on an untrained network in Section \ref{sec:mirrba}.

\subsection{Deep Image Prior}
\label{sec:dip}

The DIP method proposed in \cite{Ulyanov2020} uses a deep architecture to denoise images using a network, but without any prior learning. Supposing $X_0$ a distorted image and $X$ the network output, the fitting process is characterized by Eq.~\ref{eq:dip}, with $\mathcal{R}_{archi}$ the implicit prior captured by the network architecture and $\mathcal{S}_{DIP}$ a reconstruction function on a single image.
\begin{equation}
    \label{eq:dip}
    \underset{X(\mathcal{R}_{archi})}\argmin \hspace{1mm} \left\{ \mathcal{S}_{DIP}(X_0, X) \right\}
\end{equation}
DIP reconstructs a noisy image (e.g. with JPEG compression noise) from a white noise image by training a generator architecture to fit the noisy image. A denoised image is obtained by stopping the training before completely fitting the noise. We propose to adapt this idea to the registration of a pair of images, where  we modify the moving image to match the fixed one. Contrary to the original DIP, we are interested in fitting all the way up to the finest deformations.

\subsection{\texttt{MIRRBA} (Medical Image Registration method Regularized by Architecture)}
\label{sec:mirrba}

In this paper, we argue that deep architectures, as parametric models with high capacity, are powerful representations for deformation fields. They can thus be exploited as implicit priors for image registration in iterative optimization schemes without a training stage (\cite{Heckel2020,Dittmer2020,Lucas2018}). Although DIP priors have been explored in the context of image reconstruction, there is no prior in depth study of such architecture priors in the context of image registration.

Since there is no learning step, the training dataset no longer influences the transformation. Thus, we transform Eq.~\ref{eq:reg_3} into Eq.~\ref{eq:reg_4} to directly optimize a patient-specific CNN for the pair of interest $(F$, $M)$, as traditionally done with iterative optimization methods (see Fig.~\ref{fig:methods}.A).
\begin{equation}
    \label{eq:reg_4}
    \underset{\Phi(\mathcal{R}_{archi})}\argmin \hspace{1mm} \{ \mathcal{S}(F, W) + \lambda_{smooth} \mathcal{R}_{smooth}(\Phi) + \lambda_{diffeo}\mathcal{R}_{diffeo}(\Phi) \}
\end{equation}
By optimizing Eq.~\ref{eq:reg_4}, we find the best warped image allowed by the over-parametrization of the architecture. Carefully designing the architecture to be used, it is possible to integrate several of the commonly used tricks in conventional approaches. In this paper, we rely on the LapIRN architecture (\cite{Mok2020}), which incorporates filtering through convolutions, pyramidal coarse-to-fine refinement, and interpolation steps with down- and transpose convolutions. Eq.~\ref{eq:reg_4} is optimized with an SGD for an input pair of images. Our idea  applies to other architectures and optimization algorithms.

\section{Experimental setup}

\subsection{Dataset description}

We ran our experiments on images from a private dataset, acquired in the context of the ongoing prospective multicentric EPICURE\textsubscript{seinmeta} study (NCT03958136) for metastatic breast cancer monitoring(\cite{Colombie2021})\footnote{ \href{https://projet-epicure.fr/}{https://projet-epicure.fr/}}.
Patients underwent between two and three PET/CT acquisitions, corresponding to pre-, early- (after a month) and mid-treatment time points.
A total number of 110 pairs of images were obtained, a pair being composed of a pre-treatment and either an early- or a mid-treatment image (58 and 52 images respectively). Images were acquired at two different centers. The 54 pairs of images from 
center A were obtained using a Philips Vereos or a GE Discovery PET/CT imaging systems, while the 56 pairs of images from center B 
were acquired on two different dual-slice Siemens Biograph PET/CT scanners. 

Since we are interested in lesion monitoring in the context of metastatic breast cancer, we worked only on PET images, as shown useful in previous studies \cite{Carlier2015} and \cite{Avril2016}. 
Expert physicians manually segmented all lesions. The brain and the bladder were also delineated, since they can be useful to mask irrelevant regions for patient response assessment.
All PET images were normalized by the standardized uptake value (SUV) (\cite{Kim1994}) and resampled to obtain an isotropic resolution of $1\times1\times1$~mm\textsuperscript{3} with 200 pixels for each side. Besides, no prior registration of any kind was performed. 

This prognostic study was approved by the French Agence Nationale de S\'ecurit\'e du M\'edicament et des produits de sant\'e (ANSM, \#2018-A00959-46) and the Comit\'e de Protection des Personnes (CPP) IDF I, Paris, France (\#CPPIDF1-2018-ND40-cat.1)
and a written informed consent was obtained from all patients.

\subsection{Architectural implementation details}
\label{sec:archi_details}

Our \texttt{MIRRBA} method relies on the LapIRN network architecture proposed in \cite{Mok2020}. LapIRN is a pyramidal network with $N=3$ depth levels, each level being composed of a feature encoder, a set of residual blocks and a feature decoder, as shown in Fig.~\ref{fig:methods}. 
For each level $L_{i \in \{1, 2, 3\}}$, input images are downsampled by a factor $0.5^{(N-1)}$ using a trilinear interpolation. Hence, for the coarsest level $L_1$ the image resolution is divided by 4, while for the finest $L_3$ it remains identical. Moreover, a scaling and squaring module (\cite{Dalca2019}) enforces diffeomorphic deformations.

Network levels were trained in a coarse-to-fine manner, meaning the coarsest level $L_1$ is first trained alone, and  then higher levels are progressively trained to refine the registration. To avoid unstable starts when training levels $L_{i>1}$, lower levels weights were frozen for a fixed number of epochs. Regarding the optimization process, the learning rate was set to $10e{-4}$, and the Adam optimizer was used for 1000 iterations on the two lower levels and 2000 iterations on the finest. As a dissimilarity metric, we used the NCC, regularized by smooth and diffeomorphic terms, for which $\lambda_{smooth}$ and $\lambda_{diffeo}$ were determined with a grid search to $0.1$ and $1.0$ respectively to minimize the overall loss function over a test set (see Eq.~\ref{eq:reg_4}), as described below for DL-based methods training. Indeed, since we aim at using registration-based features to monitor metastatic breast cancer in future work, we need to reach precise global and local registration. More details can be found in \cite{Mok2020}.

Starting from the architecture in Fig.~\ref{fig:methods}, we performed an ablation study to measure the influence of each component.
Hence, we i) changed the depth of the network (network with 1, 2, 3 or 4 depth levels), ii) computed the results after training each level during the coarse-to-fine registration i.e. the coarsest, the intermediate and the finest, iii) deleted the residual connections of residual blocks to transform them into simple convolutional blocks, iv) replaced the down- and up-convolutions to respectively max-pooling and upsampling operations, v) used deformable convolutions (\cite{Dai2017}) in the finest level, and vi) used a Gaussian noise image $\mathcal{N}(0, 0.001)$ (as in \cite{Laves2019}) or concatenated fixed and moving images as network inputs. Moreover, we also vii) set both $\lambda_{smooth}$ and $\lambda_{diffeo}$ to $0$ to remove the registration-specific regularization terms (see Fig.~\ref{fig:methods}.C). All these implementations are detailed in Table \ref{tab:methods}.

\begin{table*}[t]
    \caption{Loss terms and structural setups of the methods presented in the paper. ``Pyr. net.'' stands for ``Pyraminal network'',  ``Sym. diffeo. transfo.'' for ``Symmetric diffeomorphic transformation'', ``Max.'' for ``Maxpooling'', ``Up.'' for ``Upsampling'', and ``Def. conv.'' for ``Deformable convolutions''. An absence of information in the \textit{Other} column indicates the use of setups described in Sections \ref{sec:archi_details} and \ref{sec:archi_ref_details}.}
    \label{tab:methods}
    \resizebox{\textwidth}{!}{
    \centering
    \begin{tabular}{@{}l|cccc|ccccc@{}}
        \toprule \toprule
        & \multicolumn{4}{c|}{Loss terms}  & \multicolumn{5}{c}{Structural choices}   \\ \midrule
        \multicolumn{1}{c|}{}  & $\mathcal{R}_{dataset}$ & $\mathcal{R}_{smooth}$ & $\mathcal{R}_{diffeo}$ & $\mathcal{R}_{archi}$ & Model & Depth & \begin{tabular}[c]{@{}c@{}}Trained\\levels\end{tabular} & \begin{tabular}[c]{@{}c@{}}Input\\images\end{tabular} & Other \\ \midrule \midrule
        \texttt{MIRRBA} &  & x & x & x & Pyr. net.& 3 & 3 & Moving & - \\ \midrule
        \texttt{DL\_LapIRN} & x & x & x & x & Pyr. net.& 3 & 3 & \begin{tabular}[c]{@{}c@{}}Fixed \&\\ moving\end{tabular} & - \\ \midrule
        \texttt{DL\_Voxelmorph} & x & x & x & x & \begin{tabular}[c]{@{}c@{}}U-shaped\\ network\end{tabular} & 1 & 1 & \begin{tabular}[c]{@{}c@{}}Fixed \&\\ moving\end{tabular} & - \\ \midrule
        \texttt{Elastix} & & x & & & \begin{tabular}[c]{@{}c@{}}Cubic\\ B-splines\end{tabular} & 4 & 4 & Moving & - \\ \midrule
        \texttt{SyN} & & x & & & \begin{tabular}[c]{@{}c@{}}Sym. diffeo.\\ transfo.\end{tabular} & 3 & 3 & Moving & - \\ \midrule
        \texttt{SyNCC} & & x & & & \begin{tabular}[c]{@{}c@{}}Sym. diffeo.\\ transfo.\end{tabular} & 3 & 3 & Moving & - \\ \midrule
        \texttt{Combined} & & x & x & x & \begin{tabular}[c]{@{}c@{}}Sym. diffeo.\\transfo. \&\\Pyr. net.\end{tabular} & 3 - 4 & 3 - 4 & \begin{tabular}[c]{@{}c@{}}Fixed \&\\ moving\end{tabular} & - \\ \midrule
        \texttt{MIRRBA\_wo\_Regu}  & & & & x & Pyr. net.& 3 & 3 & Moving & - \\ \midrule
        \texttt{MIRRBA\_wo\_Archi} & & x & x & & Pyr. net.& 3 & 3 & Moving & - \\ \midrule
        \texttt{MIRRBA\_Depth\_1} & & x & x & x & Pyr. net.& 1 & 1 & Moving & - \\ \midrule
        \texttt{MIRRBA\_Depth\_2} & & x & x & x & Pyr. net.& 2 & 2 & Moving & - \\ \midrule
        \texttt{MIRRBA\_Depth\_4} & & x & x & x & Pyr. net.& 4 & 4 & Moving & - \\ \midrule
        \texttt{MIRRBA\_Level\_1} & & x & x  & x & Pyr. net.& 3 & 1 & Moving & - \\ \midrule
        \texttt{MIRRBA\_Level\_2} &  & x & x& x & Pyr. net.& 3 & 2 & Moving & - \\ \midrule
        \texttt{MIRRBA\_Max} & & x & x & x & Pyr. net.& 3 & 3 & Moving & Max. \\ \midrule
        \texttt{MIRRBA\_Up} & & x & x & x & Pyr. net.& 3 & 3 & Moving & Up. \\ \midrule
        \texttt{MIRRBA\_DefConv} & & x & x & x & Pyr. net.& 3 & 3 & Moving & Def. conv.\\ \midrule
        \texttt{MIRRBA\_NoiseImg} & & x & x & x & Pyr. net.& 3 & 3 & \begin{tabular}[c]{@{}c@{}}White\\ noise\end{tabular} & - \\ \midrule
        \texttt{MIRRBA\_FixImg} & & x & x & x & Pyr. net.& 3 & 3 & \begin{tabular}[c]{@{}c@{}}Fixed \&\\ moving\end{tabular} & - \\ \midrule
        \texttt{MIRRBA\_Best} & & x & x & x & Pyr. net.& 4 & 4 & \begin{tabular}[c]{@{}c@{}}Fixed \&\\moving\end{tabular} & \begin{tabular}[c]{@{}c@{}}Max. Up.\\ Def. conv.\end{tabular} \\ \bottomrule \bottomrule
\end{tabular}}
\end{table*}

\subsection{Reference methods implementation details}
\label{sec:archi_ref_details}

Regarding the methods used as reference, we used a grid search on a test set (as described below for DL-based methods training) to find hyperparameters yielding best performance running the ANTs pipeline (\cite{Avants2009}), i.e. with a three resolutions coarse-to-fine optimization, a gradient step of 0.2 and a symmetry transformation penalty. We used both MI and NCC as similarity metrics, leading to \texttt{SyN} and \texttt{SyNCC} methods respectively. We also ran the Elastix pipeline (\cite{Klein2010}) to perform successive rigid, affine and deformable image registration with four resolutions optimized with an adaptive SGD minimizing the NCC similarity term for 1000 iterations (\texttt{Elastix}). As for our method, we used a penalty on the bending energy as regularization to enforce smooth deformations.

To highlight the regularizing power of the architecture in \texttt{MIRRBA}, we ran the pipeline without the network, optimizing directly the deformation field (initialized as a Gaussian noise $\mathcal{N}(0, 0.001)$) with the Adam optimizer, \texttt{MIRRBA\_wo\_Archi} (see Table \ref{tab:methods} and Fig~\ref{fig:methods}.D).

Regarding deep learning-based methods, \texttt{DL\_LapIRN} (\cite{Mok2020}) was ran with a similar regularization balance as \texttt{MIRRBA}, i.e. with $\lambda_{smooth}=0.1$, while all other recommended settings were used. For \texttt{DL\_Voxelmorph}, we used the latest diffeomorphic version to date (\cite{Dalca2019}), with the NCC loss and recommended settings. For methods relying on a training stage, we split our dataset into five folds, paying attention to balance data from different acquisition centers among folds. For each fold, we refer to $\mathcal{D}_{\rm train}$ and $\mathcal{D}_{\rm test}$ as the train and test dataset, respectively. We trained the DL-based approaches on $\mathcal{D}_{\rm train}$ before testing them on $\mathcal{D}_{\rm test}$ for the five folds. 

All architectures were implemented with PyTorch (\cite{Paszke2017}) and trained from scratch on a Nvidia V100 32GB SXM2 GPU.

\subsection{Evaluation metrics}
\label{sec:eval_metric}

The first criterion we used to evaluate the registration accuracy is the \textit{detection rate}, defined for an individual as the percentage of lesions presenting an overlap greater than 50\% between the wrapped and fixed segmentations (\cite{Moreau2020}). A higher percentage refers to a better detection.

To evaluate the registration more locally, we computed \textit{Dice scores} between fixed and wrapped i) brain and bladder and ii) lesions segmentation masks. A Dice score close to 1 means a precise local registration, while a Dice close to 0 is unsatisfactory. Since the detection rate represents the percentage of lesions correctly detected, it is positively correlated with the Dice score of the lesions. 
However, since some lesions (155) are cured over time and disappear on PET images, we removed them from the Dice score and detection rate computations to avoid erroneous null values. Instead, we evaluated the capacity of a method to effectively make lesions disappear by computing its \textit{disappearing rate}, or percentage of volume reduction of a lesion induced by the registration, where a complete disappearance would mean a rate of 100\%. 

Registration smoothness was evaluated by measuring, for every deformation field, the standard deviation of its Jacobian determinant \textit{SDJDet}. Null values indicate an identity transformation and high ones disorganized and incoherent displacements. Although an optimal value is difficult to define, we sought to obtain small positive values, characterizing smooth deformations (\cite{Mok2020}).

Finally, we evaluated the approximate running time of each approach, using a CPU for \texttt{SyN}, \texttt{SyNCC} and \texttt{Elastix}, and a GPU for all other methods.

\subsection{Statistical analysis}

To evaluate the statistical significance of our results, we first studied their distribution. According to the Shapiro-Wilk test (testing the null hypothesis that a sample comes from a normal distribution), we cannot reject the null hypothesis. 
Hence, we ran a paired sample t-test on our results and considered them of statistical significance if $p < 0.05$.

\section{Results}

\begin{figure*}[t]
     \centering
     \includegraphics[width=\textwidth]{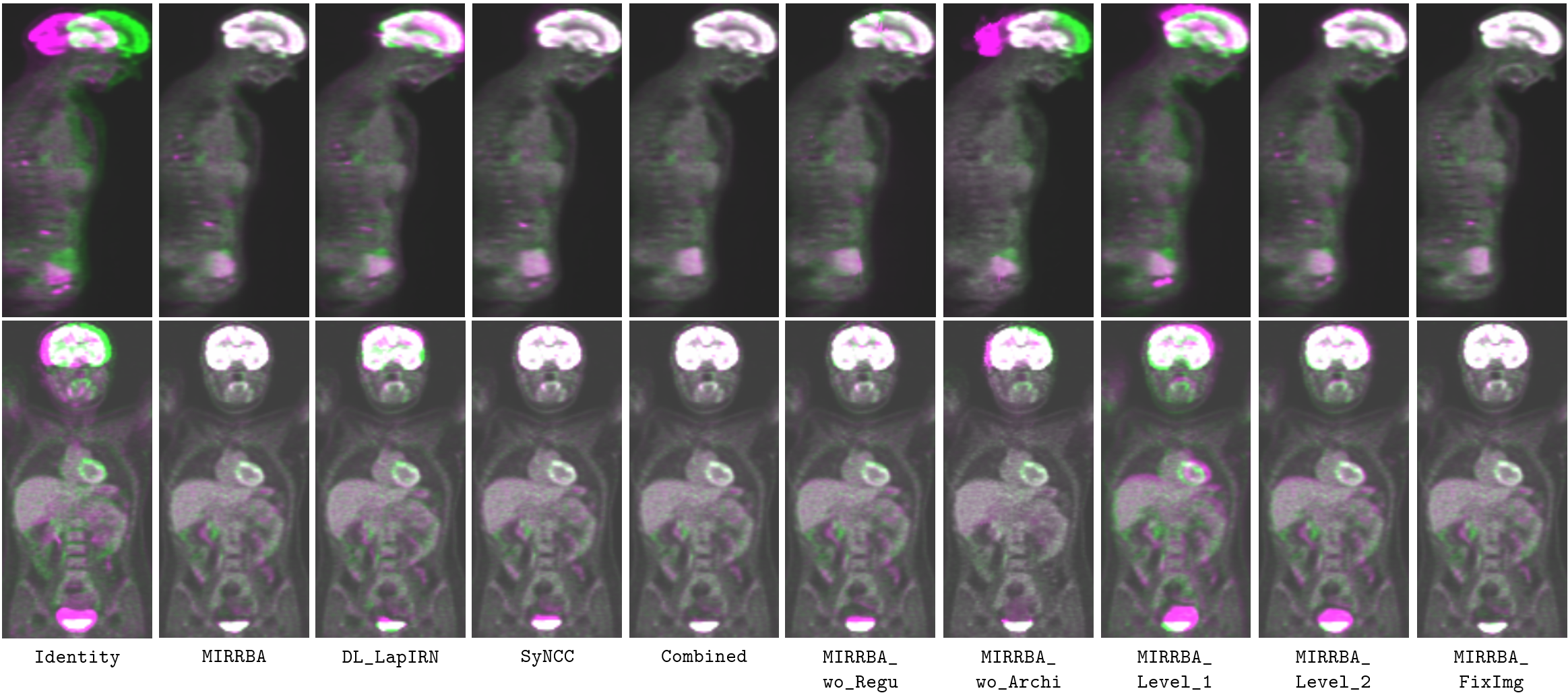}
     \caption{Overlay of the fixed (green) and warped (pink) images on two different patients after performing \texttt{Identity} registration, \texttt{MIRRBA} (corresponding to level 3), \texttt{DL\_LapIRN}, \texttt{SyNCC}, \texttt{Combined}, \texttt{MIRRBA\_wo\_Regu}, \texttt{MIRRBA\_wo\_Archi}, \texttt{MIRRBA\_Level\_1}, \texttt{MIRRBA\_Level\_2} and \texttt{MIRRBA\_FixImg}. Grayscale color indicates good overlapping. It can be noticed that \texttt{MIRRBA} registration is better than \texttt{DL\_LapIRN} and \texttt{SyNCC} around the bladder, while \texttt{Combined} obtains nicely registered images. Even without registration-specific regularization terms, \texttt{MIRRBA\_wo\_regu} wrapped images are realistic looking, unlike \texttt{MIRRBA\_wo\_archi}. We can also note that the coarsest level (\texttt{MIRRBA\_Level\_1}) performs global registration, while finest ones refine it  (\texttt{MIRRBA\_Level\_2}) and achieve more local registration (\texttt{MIRRBA}). \texttt{MIRRBA\_FixImg} reaches visually accurate registration. Best viewed in color.}
     \label{fig:01N-008_level_regu}
\end{figure*}

\subsection{Regularization terms}
\label{sec:results_regu}

In the first experiment, we looked at the influence of the different regularization terms from Eq. \ref{eq:reg_3}, i.e. $\mathcal{R}_{smooth}$, $\mathcal{R}_{diffeo}$, $\mathcal{R}_{dataset}$ and $\mathcal{R}_{archi}$. To this end, we compared the results to our \texttt{MIRRBA} approach optimized directly (without learning) on the $\mathcal{D}_{\rm test}$ pairs, using $\mathcal{R}_{smooth}$, $\mathcal{R}_{diffeo}$ and $\mathcal{R}_{archi}$ terms (see Table \ref{tab:methods}, Fig. \ref{fig:methods}.A and Eq.~\ref{eq:reg_4}).

\paragraph{Dataset regularization}
\label{sec:results_DL}

We trained the Voxelmorph (\texttt{DL\_Voxelmorph}) (\cite{Dalca2019}) and LapIRN (\texttt{DL\_LapIRN}) (\cite{Mok2020}) with the four regularization terms from Eq.~\ref{eq:reg_3} (see Table \ref{tab:methods} and Fig.~\ref{fig:methods}.B). Predictions were made on $\mathcal{D}_{\rm test}$ images.
Even if both DL-based methods reached similar numerical results (see Table~\ref{tab:results_dl_methods}) and coherent visual ones (see Fig.~\ref{fig:01N-008_level_regu} and supplementary material), \texttt{MIRRBA} performed better than both of them globally and locally. Indeed, from Table \ref{tab:results_dl_methods}, we note that running a patient-specific optimization, i.e. removing $\mathcal{R}_{dataset}$ and optimizing an untrained network \texttt{MIRRBA} compared to \texttt{DL\_Voxelmorph} and \texttt{DL\_LapIRN}, helps improve the results especially for the lesions. The organs' Dice score improved by 6\% and 5\% while the Dice of the lesions improved by 52\% and 65\% for \texttt{DL\_Voxelmorph} and \texttt{DL\_LapIRN} respectively. Moreover, \texttt{MIRRBA} presented lower SDJDet values, hence produced smoother deformations than training-based approaches. On the other hand, \texttt{DL\_Voxelmorph} and \texttt{DL\_LapIRN} had a higher disappearing rate.

\paragraph{Registration-specific regularization}

To understand the influence of the registration-specific regularization terms $\mathcal{R}_{smooth}$ and $\mathcal{R}_{diffeo}$, we removed them from Eq.~\ref{eq:reg_4} in \texttt{MIRRBA\_wo\_Regu} (see Table~\ref{tab:methods}) and Fig.~\ref{fig:methods}.C).
As quantitatively show in Table \ref{tab:results_config}, \texttt{MIRRBA} presented Dice scores for the organs and lesions which are respectively 6\% and 4\% higher compared to \texttt{MIRRBA\_wo\_Regu}. Yet, the detection and disappearing rates were increased for \texttt{MIRRBA\_wo\_Regu} over \texttt{MIRRBA}. 

\paragraph{Architecture regularization}

Finally, to also understand the impact of the regularization power of the network architecture on the registration, we looked at the results of \texttt{MIRRBA\_wo\_Archi}, which directly optimizes the deformation field without including any specific architecture, hence not including $\mathcal{R}_{archi}$ (see Table~\ref{tab:methods} and Fig.~\ref{fig:methods}.D).
According to Table \ref{tab:results_config}, \texttt{MIRRBA\_wo\_Archi} was outperformed by \texttt{MIRRBA} for both organ and lesion segmentations by 22\% and 78\% respectively. Although \texttt{MIRRBA\_wo\_Archi} disappearing rate was higher than \texttt{MIRRBA}'s, the method without architectural regularization presented a low detection rate confirming the structural bias of convolutional generators.

\begin{table*}[t]
    \caption{Comparison of \texttt{MIRRBA} to DL-based methods (Section \ref{sec:results_DL}) -- Detection rate, Dice score of the organs and lesions (Dice\_organs and Dice\_lesions resp.), SDJDet, disappearing rate, and approximate computational time. Both training and inference computational times are indicated for DL-based methods. All pipelines were computed on GPU. Statistically significant improvement of our \texttt{MIRRBA} method over the others with $p<0.05$ is indicated with~*. Best results are marked in bold, and second best ones in bold--italic, except for SDJDet since no ideal value is defined.}
    \label{tab:results_dl_methods}
    \resizebox{\textwidth}{!}{
    \centering
    \begin{tabular}{@{}lcccccc@{}}
        \toprule
         & \begin{tabular}[c]{@{}c@{}}Detec.\\rate (\%) $\uparrow$\end{tabular} & \begin{tabular}[c]{@{}c@{}}Dice\\organs $\uparrow$\end{tabular} & \begin{tabular}[c]{@{}c@{}}Dice\\lesions $\uparrow$\end{tabular} & SDJDet $\downarrow$ & \begin{tabular}[c]{@{}c@{}}Disap.\\ rate (\%) $\uparrow$\end{tabular} & \begin{tabular}[c]{@{}c@{}}Time\\(min) $\downarrow$\end{tabular} \\ \midrule
        \texttt{Identity} & 5.00 & 0.626 ± 0.138* & 0.090 ± 0.115* & 0.000 ± 0.000 & 0.00* & 0 \\ \midrule
        \texttt{MIRRBA} & \textbf{33.04} & \textbf{0.918 ± 0.126} & \textbf{0.425 ± 0.207} & 0.124 ± 0.988 & \textit{\textbf{9.36}} & 55 \\ \midrule
        \texttt{DL\_LapIRN} & 11.76 & \textbf{\textit{0.878 ± 0.076}}* & 0.258 ± 0.198* & 0.464 ± 1.367 & \textbf{19.15} & 1450 -- 3 \\
        \texttt{DL\_Voxelmorph} & \textbf{\textit{14.13}} & 0.865 ± 0.077* & \textbf{\textit{0.279 ± 0.192}}* & 0.224 ± 0.186 & 5.67 & 1200 -- 2 \\
        \bottomrule
    \end{tabular}}
\end{table*}

\begin{table*}[t]
    \caption{Ablation study on the regularization terms, architectural choices and inputs (Sections \ref{sec:results_regu} \& \ref{sec:results_ablation}) -- Detection rate, Dice score of the organs and lesions, SDJDet, disappearing rate, and approximate computational time. All pipelines were computed on GPU. Statistically significant improvement of our \texttt{MIRRBA} method over the others with $p<0.05$ is indicated with~*. Best results are marked in bold, and second best ones in bold--italic, except for SDJDet since no ideal value is defined.}
    \label{tab:results_config}
    \resizebox{\textwidth}{!}{
    \centering
    \begin{tabular}{@{}lcccccc@{}}
        \toprule
         & \begin{tabular}[c]{@{}c@{}}Detec.\\rate (\%) $\uparrow$\end{tabular} & \begin{tabular}[c]{@{}c@{}}Dice\\organs $\uparrow$\end{tabular} & \begin{tabular}[c]{@{}c@{}}Dice\\lesions $\uparrow$\end{tabular} & SDJDet $\downarrow$ & \begin{tabular}[c]{@{}c@{}}Disap.\\ rate (\%) $\uparrow$\end{tabular} & \begin{tabular}[c]{@{}c@{}}Time\\(min) $\downarrow$\end{tabular} \\ \midrule
        \texttt{Identity} & 5.00 & 0.626 ± 0.138* & 0.090 ± 0.115* & 0.000 ± 0.000 & 0.00* & 0 \\ \midrule
        \texttt{MIRRBA} & 33.04 & 0.918 ± 0.126 & 0.425 ± 0.207 & 0.124 ± 0.988 & 9.36 & 55 \\ \midrule
        \texttt{MIRRBA\_wo\_Regu} & 33.70 & 0.868 ± 0.199* & 0.407 ± 0.219* & 6.655 ± 30.709 & 16.84 & 55 \\
        \texttt{MIRRBA\_wo\_Archi} & 20.54 & 0.753 ± 0.144* & 0.239 ± 0.223* & 1.247 ± 0.475 & \textbf{59.95} & 20 \\ \midrule
        \texttt{MIRRBA\_Depth\_1} & 11.74 & 0.748 ± 0.169* & 0.239 ± 0.211* & 0.016 ± 0.047 & 4.82 & 30 \\
        \texttt{MIRRBA\_Depth\_2} & 23.26 & 0.873 ± 0.135* & 0.364 ± 0.211* & 0.038 ± 0.174 & 7.56 & 45 \\
        \texttt{MIRRBA\_Depth\_4} & \textit{\textbf{40.22}} & \textbf{\textit{0.945 ± 0.012}} & \textbf{\textit{0.466 ± 0.199}} & 0.057 ± 0.058 & 15.12 & 60 \\ \midrule
        \texttt{MIRRBA\_Level\_1} & 4.35 & 0.722 ± 0.091* & 0.221 ± 0.157* & 0.013 ± 0.010 & 0.00* & \textbf{2} \\
        \texttt{MIRRBA\_Level\_2} & 19.78 & 0.869 ± 0.114* & 0.371 ± 0.198* & 0.035 ± 0.1737 & 0.00* & \textbf{\textit{10}} \\ \midrule
        \texttt{MIRRBA\_Max} & 33.26 & 0.922 ± 0.098 & 0.426 ± 0.213 & 0.316 ± 2.934 & 8.34 & 55 \\
        \texttt{MIRRBA\_Up} & 33.04 & 0.922 ± 0.112 & 0.439 ± 0.204* & 0.056 ± 0.309 & 8.84 & 55 \\
        \texttt{MIRRBA\_DefConv} & 35.22 & 0.487 ± 0.463* & 0.257 ± 0.275* & 0.027 ± 0.020 & 12.23 & 130 \\ \midrule
        \texttt{MIRRBA\_NoiseImg} & 22.44 & 0.892 ± 0.081* & 0.381 ± 0.194* & 0.007 ± 0.009 & 0.00* & 55 \\
        \texttt{MIRRBA\_FixImg} & 36.03 & 0.941 ± 0.017 & 0.451 ± 0.197 & 0.071 ± 0.095 & \textit{\textbf{20.52}} & 60 \\ \midrule
        \texttt{MIRRBA\_Best} & \textbf{40.94} & \textbf{0.947 ± 0.010} & \textbf{0.467 ± 0.202} & 0.080 ± 0.085 & 19.48 & 60 \\ 
        \bottomrule
    \end{tabular}}
\end{table*}

\subsection{Architectural choices}
\label{sec:results_ablation}

To study the regularization effect of a network architecture on registration, we compared the results of various architectural choices built from the pyramidal network presented in \cite{Mok2020} (see Table \ref{tab:methods}). Quantitative results are presented in Table \ref{tab:results_config}, while qualitative ones are visible Fig.~\ref{fig:01N-008_level_regu} and in supplementary materials. 

\paragraph{Depth of the pyramidal network}

First, we modified the architecture to optimize a simple U-Net-shaped network (\texttt{MIRRBA\_Depth\_1}), and pyramidal ones with two (\texttt{MIRRBA\_Depth\_2}), three (\texttt{MIRRBA}), and four (\texttt{MIRRBA\_Depth\_4}) resolutions. While \texttt{MIRRBA\_Depth\_1} was optimized on full resolution images, all other architectures were trained using a coarse-to-fine strategy.

Results show that increasing the network depth improves the Dice results, as well as the detection and disappearing rates. Indeed, \texttt{MIRRBA\_Depth\_4} presented the second highest Dice scores for both the organs and the lesions among all MIRRBA setups. Regarding the SDJDet values, they were higher when using a depth 3 or 4, than when the network was trained on less resolutions. 

\paragraph{Trained network level}

To understand the amount of information brought by each network level during the coarse-to-fine training strategy, we computed the registration after optimizing only the lower level (\texttt{MIRRBA\_Level\_1}) on coarse resolution images, both lower levels (\texttt{MIRRBA\_Level\_2}) on coarse and medium resolution images, and the whole network (\texttt{MIRRBA}) with the complete coarse-to-fine training strategy.

According to Table \ref{tab:results_config}, training on all three levels of the network improved the registration accuracy over training only on low resolution images. Moreover, the disappearing rate was null when the training only occurred on low resolution images.

\paragraph{Max-pooling and upsampling operations}

For each network level, we replaced the down-convolution (convolution with stride 2) by a max-pooling operation in \texttt{MIRRBA\_Max}, and the transpose convolution by an upsampling in \texttt{MIRRBA\_Up}. 
As presented Table \ref{tab:results_config}, using these learning-free operations improved the registration accuracy globally and locally. The disappearing rate was however slightly reduced.

\paragraph{Residual blocks}

We removed the residual connections of the residual blocks (\texttt{MIRRBA\_wo\_RB}) to understand their influence in our architecture.
Without them, we obtained null Dice scores and detection rate, as well as very high SDJDet value. Besides, wrapped images did not look realistic (see Fig.~\ref{fig:01N-008_01A-016_network} in the supplementary materials).

\paragraph{Deformable convolutions}

To adapt the receptive field of the convolutions to the local scale of objects to be registered, we replaced those of the highest resolution level (i.e. level 3) by deformable convolutions (\texttt{MIRRBA\_DefConv}). While global registration results were improved over \texttt{MIRRBA}, local ones were not. Moreover, the SDJDet was low.

\paragraph{Input images}

Regarding the network inputs, instead of conditioning the network with the moving image, we fed it with a Gaussian noise, as in \cite{Laves2019}. In Table \ref{tab:results_config}, \texttt{MIRRBA\_NoiseImg} shows worse results than \texttt{MIRRBA}, except for the SDJDet. 
We also provided more information to the network by concatenating the fixed to the moving image in \texttt{MIRRBA\_FixImg}, which significantly improved all \texttt{MIRRBA} results, especially the disappearing rate.

\paragraph{Combining best practices}

Finally, we combined the best architectural variations presented above to perform registration with four resolutions, both fixed and moving images as input, max-pooling and upsampling operations, as well as residual blocks: \texttt{MIRRBA\_Best}.

\texttt{MIRRBA\_Depth\_4} Dice scores were slightly improved by the use of max-pooling and upsampling operations. On the other hand, adding a fourth depth to \texttt{MIRRBA\_FixImg}, \texttt{MIRRBA\_Max} and \texttt{MIRRBA\_Up} statistically improved their results, leading to the best performing MIRRBA-based method in terms of local registration precision \texttt{MIRRBA\_Best}. The SDJDet of \texttt{MIRRBA\_Best} was smaller or similar to either one of the three other methods, while the disappearing rate lied between their values. Images of lesions registered with \texttt{MIRRBA\_Best} are visible Fig.~\ref{fig:lesions}.

\subsection{Comparison to conventional methods}
\label{sec:results_conv}

\paragraph{Conventional registration}

\begin{table*}[t]
    \caption{Comparison of \texttt{MIRRBA} to conventional methods (Section \ref{sec:results_conv}) -- Detection rate, Dice score of the organs and lesions (Dice\_organs and Dice\_lesions resp.), SDJDet, disappearing rate, and approximate computational time. While ANTs and Elastix pipeline were computed on CPU, \texttt{MIRRBA} ran on GPU. Statistically significant improvement of our \texttt{MIRRBA} method over the others with $p<0.05$ is indicated with~*. Best results are marked in bold, and second best ones in bold--italic, except for SDJDet since no ideal value is defined.}
    \label{tab:results_conv_methods}
    \resizebox{\textwidth}{!}{
    \centering
    \begin{tabular}{@{}lcccccc@{}}
        \toprule
         & \begin{tabular}[c]{@{}c@{}}Detec.\\rate (\%) $\uparrow$\end{tabular} & \begin{tabular}[c]{@{}c@{}}Dice\\organs $\uparrow$\end{tabular} & \begin{tabular}[c]{@{}c@{}}Dice\\lesions $\uparrow$\end{tabular} & SDJDet $\downarrow$ & \begin{tabular}[c]{@{}c@{}}Disap.\\ rate (\%) $\uparrow$\end{tabular} & \begin{tabular}[c]{@{}c@{}}Time\\(min) $\downarrow$\end{tabular} \\ \midrule
        \texttt{Identity} & 5.00 & 0.626 ± 0.138* & 0.090 ± 0.115* & 0.000 ± 0.000 & 0.00* & 0 \\ \midrule
        \texttt{MIRRBA} & 33.04 & 0.918 ± 0.126 & 0.425 ± 0.207& 0.124 ± 0.988 & 9.36 & \textbf{\textit{55}} \\ \midrule
        \texttt{Elastix} & 20.54 & 0.868 ± 0.124* & 0.350 ± 0.191* & 0.096 ± 0.044 & 9.59 & 25 \\
        \texttt{SyN} & 24.57 & 0.936 ± 0.023 & 0.386 ± 0.210* & 0.016 ± 0.018 & 0.00 & \textbf{5} \\
        \texttt{SyNNC} & \textit{\textbf{39.57}} & \textit{\textbf{0.944 ± 0.014}} & \textit{\textbf{0.477 ± 0.211}} & 0.073 ± 0.066 & 4.26* & 60 \\ \midrule
        \texttt{Combined} & \textbf{44.71} & \textbf{0.945 ± 0.012} & \textbf{0.481 ± 0.197} & 0.077 ± 0.072 & \textbf{25.11} & 115 \\
        \bottomrule
    \end{tabular}}
\end{table*}

According to Table \ref{tab:results_conv_methods}, even after rigid and affine pre-registration, locally precise deformable registration is challenging on whole-body images for the conventional \texttt{Elastix} (\cite{Klein2010}) pipeline (see Fig.~\ref{fig:01N-008_01A-016_ref_methods} in supplementary material).

On the other hand, both \texttt{SyN} and \texttt{SyNCC} (\cite{Avants2009}) statistically performed better than \texttt{MIRRBA} (which used the NCC metric) on organ segmentation, while the NCC similarity metric allowed \texttt{SyNCC} to also reach a better accuracy on lesion segmentation. Regarding the disappearing rate, \texttt{MIRRBA} performed better than both SyN-based methods, whereas their SDJDet was lower than ours.

\paragraph{Combination of DIP and conventional registration}

Finally, we pushed the analysis by combining our best MIRRBA-based method with the best performing conventional \texttt{SyNCC} approach. To do so, we optimized \texttt{MIRRBA\_Best}, using as input the deformation fields and already registered images obtained by \texttt{SyNCC}. Our assumption was that the conventional pre-registration would be improved by our method. 

Results of this \texttt{Combined} pipeline are shown in Table~\ref{tab:results_conv_methods}. In terms of global and local accuracy, the combined approach outperformed both \texttt{SyNCC} and \texttt{MIRRBA\_Best}. In addition, the disappearing rate was significantly improved for both methods.

\section{Discussion}

\subsection{Impact of training on a database}

When comparing \texttt{MIRRBA} to the DL-based approaches
(\texttt{DL\_LapIRN} and \texttt{DL\_Voxelmorph}), we show that not learning registration patterns from a dataset helps to obtain precise segmentations, especially at a lesion level. Indeed, while the size of organs and their locations are relatively consistent across patients, this is not the case with the lesions. Hence, there are fewer deformation patterns that can be learned from a database. Since performing locally precise registration with a DL-based method is very challenging in this situation, our method adapts to each subject. Moreover, according to its lower SDJDet value, \texttt{MIRRBA} produces smoother deformations than the DL-based approaches.

\subsection{Architectural choices}

\paragraph{Regularization terms}
As presented in Section \ref{sec:img_reg}, removing the registration-specific regularization terms is equivalent to solving an ill-posed problem in conventional registration methods. In our case, since we use a network to perform registration, $\mathcal{R}_{archi}$ is present when running \texttt{MIRRBA\_wo\_Regu}. For similar Dice scores and detection rate, \texttt{MIRRBA\_wo\_Regu} has higher SDJDet than \texttt{MIRRBA}, indicating $\mathcal{R}_{smooth}$ and $\mathcal{R}_{diffeo}$ help smooth the deformation field, even though they have less impact than the regularization of the architecture.

On the opposite configuration \texttt{MIRRBA\_wo\_Archi}, where $\mathcal{R}_{archi}$ is not used but $\mathcal{R}_{smooth}$ and $\mathcal{R}_{diffeo}$ are, quantitative results show that the lack of architecture negatively impacts the registration smoothness, as well as its local and global precision, making questionable the convergence of the method without architecture. Hence, we confirm that the architecture has a regularization effect on the registration, which helps to find an admissible solution.

\paragraph{Coarse to fine strategies and image resolution}
Considering the overall network, \cite{Dittmer2020} affirms that an architecture ran in a DIP pipeline acts as a low-pass filter in the beginning of the optimization, allowing higher frequencies to pass only after lower ones. We observe similar results over the coarse-to-fine training strategy (see Fig.~\ref{fig:01N-008_level_regu}), where low frequencies are registered first (global registration), followed later by higher frequencies (local registration). In the same way, looking at the intermediate results during the 4-depth pyramidal optimization of \texttt{Elastix} (see Fig.~\ref{fig:01N-008_01A-016_ref_methods} in supplementary materials), we observe a tendency to register global features before local ones. 

Modifying the number of coarse resolution levels, i.e. the depth of the pyramidal architecture, we studied the impact of global structural choices to optimize the whole network. The higher SDJDet values obtained when more resolutions are used could be explained by the generation of more local transformations, hence a globally less regular deformation field.
Moreover, with four resolutions, the receptive field of the coarsest level of \texttt{MIRRBA\_Depth\_4} captures the whole image (i.e. $200~\times~200~\times~200$), explaining the high Dice scores, as well as detection and disappearing rates. Indeed, successful conventional pipelines as Elastix (\cite{Klein2010}) or ANTs (\cite{Avants2009}) also uses this type of pyramidal strategy. 

\paragraph{Network conditioning with input images}
Contrary to \cite{Ulyanov2020} and \cite{Laves2019}, who used Gaussian noise as input for their architectures, \cite{Gong2019} and \cite{Baguer2020} respectively improved CT and PET DIP-based image reconstructions by providing acquisitions from other modalities to their networks. We made similar observations, as \texttt{MIRRBA\_NoiseImg} reached less accurate results than when we conditioned the model with PET images, as in \texttt{MIRRBA}. Moreover, feeding the network with more patient information, using the fixed image as additional input to the moving image in \texttt{MIRRBA\_FixImg}, improves the results by increasing the network conditioning on a single patient.

\paragraph{Other architectural choices}
Moving inside the architecture, residual blocks can be related to diffeomorphic registration according to \cite{Rousseau2020}. Indeed, stacking residual blocks in ResNets (\cite{He2016}) aims to incrementally map the embedding space to a new unknown space, each block being defined as $y=F(x)+x$, with $x$ and $y$ the respective input and output of the residual blocks, and $F$ the residual mapping to be learned. Similarly, diffeomorphic registration models (\cite{Beg2005,Sotiras2013}) address the registration issue by piling up incremental diffeomorphic mappings. Making the link between ResNets and registration, the function $F$ can be seen as a parametrization of an elementary deformation flow, and training a series of residual blocks as learning continuous and integral diffeomorphic operator. With our deep architecture, \texttt{MIRRBA\_wo\_RB} results indicate that the registration without residual blocks fails to converge. 
Indeed, as explained above, residual blocks allow incremental diffeomorphic mappings, and removing them leads to gradient vanishing for our patient-specific method.

\begin{figure}[t]
 \centering
 \includegraphics[width=11cm]{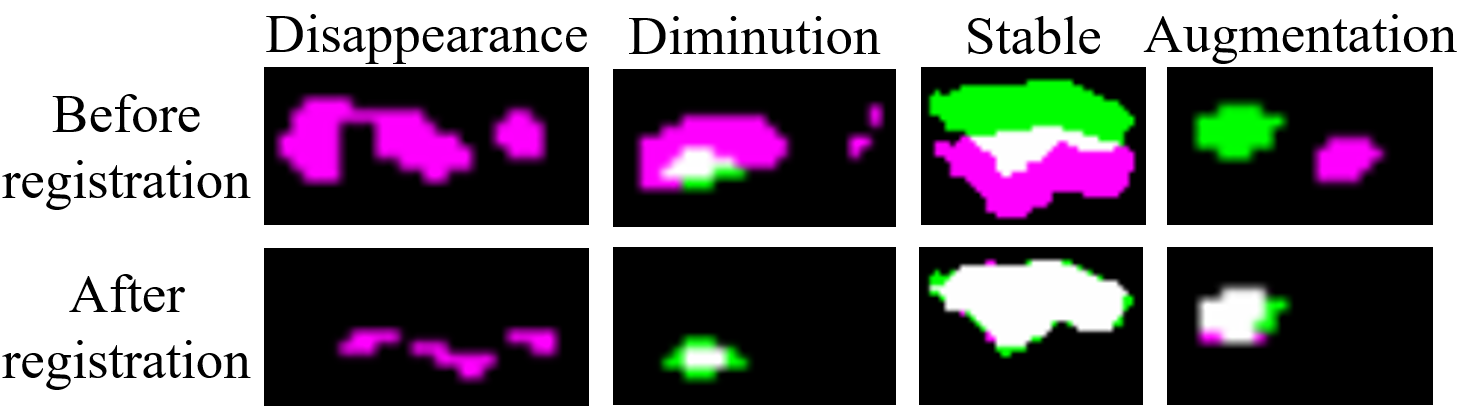}
 \caption{Overlay of lesions disappearing (\nth{1} column), reducing (\nth{2} column), stable (\nth{3} column) and growing (last column) before and after performing \texttt{MIRRBA\_Best}. Fixed lesions can be viewed in green, while moving and wrapped ones in pink. White color indicates overlapping. Best viewed in color.}
 \label{fig:lesions}
\end{figure}

Regarding the convolutions, \cite{Heinrich2019} suggested deformable convolutions (\cite{Dai2017}) to capture larger deformations. These convolutions add 2D or 3D offsets to the regular grid sampling of standard operations. If these offsets are set to zero, deformable convolutions become standard convolutions, otherwise they modify the receptive field. Since these offsets are learnable, deformable convolutions are trained to adapt their receptive fields, in order to focus on objects of interest in classification problems for instance. 
According to the high detection rate, this is performed globally by \texttt{MIRRBA\_DefConv}. Yet, the resulting low Dice scores show that \texttt{MIRRBA\_DefConv} did not achieve precise local registration, which can be explained by the low SDJDet value, showing that only smooth and regular deformations occurred instead of locally irregular ones. Indeed,  deformable convolutions might need a more complex integration in the architecture (\cite{Liu2020}). Moreover, the additional learnable parameters and adaptive receptive field of deformable convolutions are to some extent redundant with the pyramidal structure, and make it harder to train. 

Considering the amount of parameters to optimize in a network, we looked at the results of \texttt{MIRRBA\_Max} and \texttt{MIRRBA\_Up} compared to \texttt{MIRRBA}. Even if the learnable down- and up-convolutions are now common to respectively increase or decrease image dimensions within a network, max-pooling and upsampling operations were originally used. As in conventional registration methods, these operations are not learnable. Hence, their results do not depend on optimization parameters, and they help to control the overfitting in a traditional trainable setup. Therefore, the good results obtained using max-pooling and upsampling operations in \texttt{MIRRBA\_Max}, \texttt{MIRRBA\_Up} and \texttt{MIRRBA\_Depth\_4\_Max\_Up} may be explained by the fewer number of parameters to fit (see Fig.~\ref{fig:01N-008_01A-016_network} in supplementary materials).

\subsection{Lesion segmentation}

Regarding lesion registration, Fig.~\ref{fig:lesions} shows different lesion evolution scenarios. We can see that the moving masks adapt to the fixed ones when the lesion does not disappear. Although the disappearing lesion from the first column was not deleted, it was reduced by the registration algorithm. Indeed, lesion disappearance implies deformations which are not smooth nor diffeomorphic, hence the good disappearing rate of \texttt{MIRRBA\_wo\_Regu} and the probable need to adapt our registration strategy to the specific problem of lesion disappearance in future work.

\section{Conclusion}

In this paper, we propose an alternative method to perform image registration using a neural network without the typical learning stage on a database. We formalize the registration problem by following the conventional approaches relying on image-based similarities and regularization terms, but also explicitly consider the dataset and architecture bias. Indeed, our study was motivated by recent work on DIP, implying that neural networks create an inductive bias when learning from a database, but also an intrinsic structural bias induced by the architecture \cite{Heckel2020}. These biases are shown sufficient to solve certain image processing tasks. Our proposition also resonates with \cite{Dittmer2020}, who suggested that deep convolutional neural networks process low-frequency information first, to later focus on the finer deformations, both desirable properties for a registration algorithm. We integrate the LapIRN network from \cite{Mok2020}, who developed an effective pyramidal architecture, tested in the DL-based set-up. Here, we further demonstrate that beyond any prior coming from the dataset learning step, the architecture design has an important effect on the registration results, acting as an implicit regularizer. Our study also shows the impact of some of the architecture components, particularly the residual blocks, and we justify this behavior by making a link with findings from \cite{Rousseau2020}. Moreover, we found that for our application, a pyramidal architecture capturing the whole image with a limited amount of parameters to optimize, as conventional registration methods, provides precise registration results.  

The architectural prior seems to be a better option than learning from data in cases where there are no positional consistency, which is the case with metastatic breast cancer lesions, which arbitrarily vary in position, size and number. Indeed, finding a set of network parameters allowing precise registration for a whole dataset is a challenging task.
Our approach enables to correctly register active organs such as the brain and the bladder, which could be used to automatically propagate annotations masking regions irrelevant for patient response evaluation. Although the Dice scores are relatively low for the lesions, we obtain good detection values and improve the disappearance rate.

Our approach makes a step in bridging conventional and DL-based methods for image registration, and provides several suitable approaches for the challenging 3D full-body longitudinal registration problem. We demonstrated the possibility to perform both global and local registration on whole body medical images using a network but without suffering from dataset bias. In future work, we would study the feasibility of extracting registration-based feature from our method to monitor lesion evolution without depending on manually performed segmentations.

\bibliographystyle{splncs04}
\bibliography{biblio}

\begin{thebibliography}{10}
\providecommand{\url}[1]{\texttt{#1}}
\providecommand{\urlprefix}{URL }
\providecommand{\doi}[1]{https://doi.org/#1}

\bibitem{Avants2009}
Avants, B.B., Tustison, N., Johnson, H.: {Advanced Normalization Tools (ANTS)}.
  Insight Journal  \textbf{2}(365),  1--35 (2009)

\bibitem{Avril2016}
Avril, S., Muzic, R.F., Plecha, D., Traughber, B.J., Vinayak, S., Avril, N.:
  {18F-FDG PET/CT for Monitoring of Treatment Response in Breast Cancer}.
  Journal of Nuclear Medicine  \textbf{57}(Supplement 1),  34S--39S (2016).
  \doi{10.2967/jnumed.115.157875}

\bibitem{Baguer2020}
Baguer, D.O., Leuschner, J., Schmidt, M.: {Computed tomography reconstruction
  using deep image prior and learned reconstruction methods}. Inverse Problems
  \textbf{36} (2020)

\bibitem{Balakrishnan2018}
Balakrishnan, G., Zhao, A., Sabuncu, M.R., Guttag, J., Dalca, A.V.:
  {VoxelMorph: A Learning Framework for Deformable Medical Image Registration}.
  IEEE Transactions on Medical Imaging  \textbf{38}(8),  1--13 (2018).
  \doi{10.1109/TMI.2019.2897538}

\bibitem{Beg2005}
Beg, M.F., Miller, M.I., Trouv{\'{e}}, A., Younes, L.: {Computing large
  deformation metric mappings via geodesic flows of diffeomorphisms}.
  International Journal of Computer Vision  \textbf{61}(2),  139--157 (2005).
  \doi{10.1023/B:VISI.0000043755.93987.aa}

\bibitem{Carlier2015}
Carlier, T., Bailly, C.: {State-Of-The-Art and Recent Advances in
  Quantification for Therapeutic Follow-Up in Oncology Using PET}. Frontiers in
  Medicine  \textbf{2} (2015). \doi{10.3389/fmed.2015.00018}

\bibitem{Chassagnon2020}
Chassagnon, G., Vakalopoulou, M., R{\'{e}}gent, A., Sahasrabudhe, M., Marini,
  R., Hoang-Thi, T.N., Dinh-Xuan, A.T., Dunogu{\'{e}}, B., Mouthon, L.,
  Paragios, N., Revel, M.P.: {Elastic Registration – driven Deep Learning for
  Longitudinal Assessment of Systemic Sclerosis}. Radiology  \textbf{298}(1),
  189--198 (2020). \doi{10.1148/radiol.2020200319}

\bibitem{Chen2021}
Chen, X., Diaz-pinto, A., Ravikumar, N., Frangi, A.F.: {Deep learning in
  medical image registration}. Progress in Biomedical Engineering
  \textbf{3}(1) (2021). \doi{10.1088/2516-1091/abd37c}

\bibitem{Christensen2007}
Christensen, G.E., Song, J.H., Lu, W., Naqa, I.E., Low, D.A.: {Tracking lung
  tissue motion and expansion/compression with inverse consistent image
  registration and spirometry}. Medical Physics  \textbf{34}(6),  2155--2163
  (2007). \doi{10.1118/1.2731029}

\bibitem{Colombie2021}
Colombi{\'{e}}, M., J{\'{e}}z{\'{e}}quel, P., Rubeaux, M., Frenel, J.S., Bigot,
  F., Seegers, V., Campone, M.: {The EPICURE study: a pilot prospective cohort
  study of heterogeneous and massive data integration in metastatic breast
  cancer patients}. BMC Cancer  \textbf{21}(1), ~1--9 (2021).
  \doi{10.1186/s12885-021-08060-8}

\bibitem{Dai2017}
Dai, J., Qi, H., Xiong, Y., Li, Y., Zhang, G., Hu, H., Wei, Y.: {Deformable
  Convolutional Networks}. Proceedings of the IEEE International Conference on
  Computer Vision pp. 764--773 (2017). \doi{10.1109/ICCV.2017.89}

\bibitem{Dalca2019}
Dalca, A.V., Balakrishnan, G., Guttag, J., Sabuncu, M.R.: {Unsupervised
  learning of probabilistic diffeomorphic registration for images and
  surfaces}. Medical Image Analysis  \textbf{57},  226--236 (2019).
  \doi{10.1016/j.media.2019.07.006}

\bibitem{Dittmer2020}
Dittmer, S., Kluth, T., Maass, P., Baguer, D.O.: {Regularization by
  architecture : A deep prior approach for inverse problems}. Journal of
  Mathematical Imaging and Vision  \textbf{62}(3),  456--470 (2020).
  \doi{10.1007/s10851-019-00923-x}

\bibitem{Eppenhof2020}
Eppenhof, K.A., Lafarge, M.W., Veta, M., Pluim, J.P.: {Progressively Trained
  Convolutional Neural Networks for Deformable Image Registration}. IEEE
  Transactions on Medical Imaging  \textbf{39}(5),  1594--1604 (2019).
  \doi{10.1109/TMI.2019.2953788}

\bibitem{Friston1995}
Friston, K.J., Ashburner, J., Frith, C.D., Poline, J.B., Heather, J.D.,
  Frackowiak, R.S.: {Spatial registration and normalization of images}. Human
  Brain Mapping  \textbf{3}(3),  165--189 (1995). \doi{10.1002/hbm.460030303}

\bibitem{Fu2020}
Fu, Y., Lei, Y., Wang, T., Curran, W.J., Liu, T., Yang, X.: {Deep learning in
  medical image registration: a review}. Physics in Medicine and Biology
  \textbf{60}(20),  1--32 (2020). \doi{10.1088/1361-6560/ab843e}

\bibitem{Gong2019}
Gong, K., Catana, C., Qi, J., Li, Q.: {PET Image Reconstruction Using Deep
  Image Prior}. IEEE Transactions on Medical Imaging  \textbf{38}(7),
  1655--1665 (2019). \doi{10.1109/TMI.2018.2888491}

\bibitem{He2016}
He, K., Zhang, X., Ren, S., Sun, J.: {Identity Mappings in Deep Residual
  Networks}. European conference on computer vision pp. 630--645 (2016)

\bibitem{Heckel2020}
Heckel, R., Soltanolkotabi, M.: {Compressive sensing with un-trained neural
  networks: Gradient descent finds the smoothest approximation}. International
  Conference on Machine Learning  (2020), \url{http://arxiv.org/abs/2005.03991}

\bibitem{Heinrich2019}
Heinrich, M.P.: {Closing the Gap Between Deep and Conventional Image
  Registration Using Probabilistic Dense Displacement Networks}. International
  Conference on Medical Image Computing and Computer-Assisted Intervention
  \textbf{October},  50--58 (2019)

\bibitem{Heinrich2020}
Heinrich, M.P., Hansen, L.: {Highly accurate and memory efficient unsupervised
  learning-based discrete CT registration using 2.5 D displacement search}.
  International Conference on Medical Image Computing and Computer-Assisted
  Intervention pp. 190--200 (2020)

\bibitem{Hsu2020}
Hsu, T.M.H.: {Automatic Longitudinal Assessment of Tumor Responses}. Ph.D.
  thesis, Massachusetts Institute of Technology (2020)

\bibitem{Jaderberg2015}
Jaderberg, M., Simonyan, K., Zisserman, A., Kavukcuoglu, K.: {Spatial
  Transformer Networks}. Advances in neural information processing systems
  \textbf{28},  2017--2025 (2015). \doi{10.5555/2969442.2969465}

\bibitem{Kim1994}
Kim, C.K., Gupta, N.C., Chandramouli, B., Alavi, A.: {Standardized uptake
  values of FDG: Body surface area correction is preferable to body weight
  correction}. Journal of Nuclear Medicine  \textbf{35}(1),  164--167 (1994)

\bibitem{Klein2009}
Klein, A., Andersson, J., Ardekani, B.A., Ashburner, J., Avants, B., Chiang,
  M.C., Christensen, G.E., Collins, D.L., Gee, J., Hellier, P., Song, J.H.,
  Jenkinson, M., Lepage, C., Rueckert, D., Thompson, P., Vercauteren, T.,
  Woods, R.P., Mann, J.J., Parsey, R.V.: {Evaluation of 14 nonlinear
  deformation algorithms applied to human brain MRI registration}. NeuroImage
  \textbf{46}(3),  786--802 (2009). \doi{10.1016/j.neuroimage.2008.12.037}

\bibitem{Klein2010}
Klein, S., Staring, M., Murphy, K., Viergever, M., Pluim, J.: {elastix: A
  Toolbox for Intensity-Based Medical Image Registration}. IEEE Transactions on
  Medical Imaging  \textbf{29}(1),  196--205 (2010).
  \doi{10.1109/TMI.2009.2035616}

\bibitem{Klein2009a}
Klein, S., Pluim, J.P., Staring, M., Viergever, M.A.: {Adaptive stochastic
  gradient descent optimisation for image registration}. International Journal
  of Computer Vision  \textbf{81}(3),  227--239 (2009).
  \doi{10.1007/s11263-008-0168-y}

\bibitem{Laves2019}
Laves, M.H., Ihler, S., Ortmaier, T.: {Deformable medical image registration
  using a randomly-initialized CNN as regularization prior}. arXiv  \textbf{1},
  ~1--6 (2019)

\bibitem{Li2021}
Li, T., Zhang, M., Qi, W., Asma, E., Qi, J.: {Motion Correction of
  Respiratory-Gated PET Images Using Deep Learning Based Image Registration
  Framework}. Physics in Medicine and Biology  \textbf{65}(15),  155003 (2021).
  \doi{10.1088/1361-6560/ab8688.Motion}

\bibitem{Liu2020}
Liu, F., Liu, D., Tian, J., Xie, X., Yang, X., Wang, K.: {Cascaded one-shot
  deformable convolutional neural networks: Developing a deep learning model
  for respiratory motion estimation in ultrasound sequences}. Medical Image
  Analysis  \textbf{65} (2020). \doi{10.1016/j.media.2020.101793}

\bibitem{Lucas2018}
Lucas, A., Iliadis, M., Molina, R.: {Using Deep Neural Networks for Inverse
  Problems in Imaging: Beyond Analytical Methods}. IEEE Signal Processing
  Magazine  \textbf{35}(1),  20--36 (2018). \doi{10.1109/msp.2017.2760358}

\bibitem{Maurer1993}
Maurer, C., Fitzpatrick, J.: {A Review of Medical Image Registration}.
  Interactive Image-Guided Neurosurgery pp. 17--44 (1993)

\bibitem{Mok2020}
Mok, T.C.W., Chung, A.C.S.: {Large Deformation Diffeomorphic Image Registration
  with Laplacian Pyramid Networks}. International Conference on Medical Image
  Computing and Computer-Assisted Intervention pp. 211--221 (2020)

\bibitem{Moreau2020}
Moreau, N., Rousseau, C., Fourcade, C., Santini, G., Ferrer, L., Lacombe, M.,
  Guillerminet, C., Campone, M., Colombie, M., Rubeaux, M., Normand, N.: {Deep
  learning approaches for bone and bone lesion segmentation on 18FDG PET/CT
  imaging in the context of metastatic breast cancer}. Proceedings of the
  Annual International Conference of the IEEE Engineering in Medicine and
  Biology Society, EMBS  \textbf{2020-July},  1532--1535 (2020).
  \doi{10.1109/EMBC44109.2020.9175904}

\bibitem{Myronenko2009}
Myronenko, A., Song, X.: {Adaptive Regularization of Ill-Posed Problems:
  Application to Non-rigid Image Registration}. arXiv preprint arXiv:0906.3323
  pp. 1--10 (2009), \url{http://arxiv.org/abs/0906.3323}

\bibitem{Necib2011}
Necib, H., Garcia, C., Wagner, A., Vanderlinden, B., Emonts, P., Hendlisz, A.,
  Flamen, P., Buvat, I.: {Detection and Characterization of Tumor Changes in
  18F-FDG PET Patient Monitoring Using Parametric Imaging}. Journal of Nuclear
  Medicine  \textbf{52}(3),  354--361 (2011). \doi{10.2967/jnumed.110.080150}

\bibitem{Paszke2017}
Paszke, A., Gross, S., Chintala, S., Chanan, G., Yang, E., DeVito, Z., Lin, Z.,
  Desmaison, A., Antiga, L., Lerer, A.: {Automatic differentiation in PyTorch}.
  31st Conference on Neural Information Processing System  (2017)

\bibitem{Ronneberger2015}
Ronneberger, O., Fischer, P., Brox, T.: {U-net: Convolutional networks for
  biomedical image segmentation}. International Conference on Medical Image
  Computing and Computer-Assisted Intervention  \textbf{9351},  234--241 (2015)

\bibitem{Rousseau2020}
Rousseau, F., Drumetz, L., Fablet, R.: {Residual Networks as Flows of
  Diffeomorphisms}. Journal of Mathematical Imaging and Vision  \textbf{62}(3),
   365--375 (2020). \doi{10.1007/s10851-019-00890-3}

\bibitem{Rueckert1999}
Rueckert, D., Sonoda, L.I., Hayes, C., Hill, D.L.G., Leach, M.O., Hawkes, D.J.:
  {Nonrigid registration using free-form deformations: Application to breast mr
  images}. IEEE Transactions on Medical Imaging  \textbf{18}(8),  712--721
  (1999). \doi{10.1109/42.796284}

\bibitem{Sotiras2013}
Sotiras, A., Davatzikos, C., Paragios, N.: {Deformable medical image
  registration: A survey}. IEEE Transactions on Medical Imaging
  \textbf{32}(7),  1153--1190 (2013). \doi{10.1109/TMI.2013.2265603}

\bibitem{Stergios2018}
Stergios, C., Mihir, S., Maria, V., Guillaume, C., Marie-Pierre, R., Stavroula,
  M., Nikos, P.: {Linear and deformable image registration with 3D
  convolutional neural networks}. Image Analysis for Moving Organ, Breast, and
  Thoracic Images pp. 13--22 (2018)

\bibitem{Ulyanov2020}
Ulyanov, D., Vedaldi, A., Lempitsky, V.: {Deep Image Prior}. International
  Journal of Computer Vision  \textbf{128}(7),  1867--1888 (2020).
  \doi{10.1007/s11263-020-01303-4}

\bibitem{Vercauteren2007a}
Vercauteren, T., Pennec, X., Perchant, A., Ayache, N.: {Non-parametric
  diffeomorphic image registration with the demons algorithm. Medical Image
  Computing and Computer-Assisted Intervention}. International Conference on
  Medical Image Computing and Computer-Assisted Intervention  \textbf{792},
  319--326 (2007)

\bibitem{Vercauteren2007}
Vercauteren, T., Pennec, X., Malis, E., Perchant, A., Ayach, N.: {Insight into
  efficient image registration techniques and the demons algorithm}. Biennial
  International Conference on Information Processing in Medical Imaging
  \textbf{July},  495--506 (2007)

\bibitem{DeVos2019}
de~Vos, B.D., Berendsen, F.F., Viergever, M.A., Sokooti, H., Staring, M.,
  I{\v{s}}gum, I.: {A deep learning framework for unsupervised affine and
  deformable image registration}. Medical Image Analysis  \textbf{52},
  128--143 (2019). \doi{10.1016/j.media.2018.11.010}

\end{thebibliography}

\newpage
\section*{Supplementary Material}

\begin{figure*}[ht]
     \setcounter{figure}{0}
     \centering
     \includegraphics[width=\textwidth]{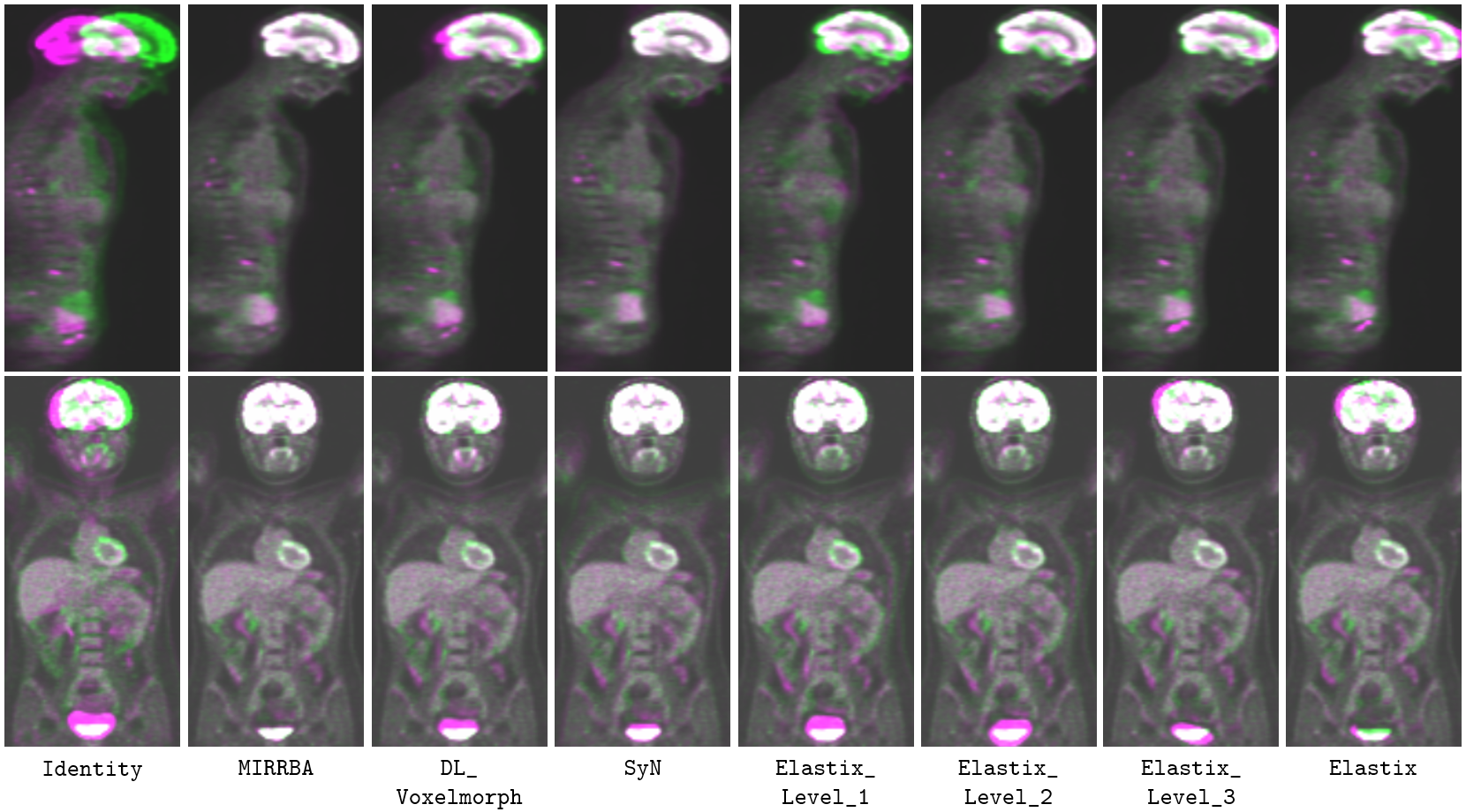}
     \caption{Comparison to reference methods -- Overlay of the fixed (green) and warped (pink) images on two different patients after performing \texttt{Identity} registration, \texttt{MIRRBA}, \texttt{DL\_Voxelmorph}, \texttt{SyN}, \texttt{Elastix\_Level\_1}, \texttt{Elastix\_Level\_2}, \texttt{Elastix\_Level\_3} and \texttt{Elastix} (corresponding to the level 4). Grayscale color indicates good overlapping. DL-based method has difficulties to register the bladder because of its important deformation. \texttt{SyN} wrapped images look coherent, even if missing a bit of precision around the bladder. The pyramidal optimization of \texttt{Elastix} acts as a progressive registration: global features are registered before more local and precise ones. Best viewed in color.}
     \label{fig:01N-008_01A-016_ref_methods}
\end{figure*}

\begin{figure*}[ht]
     \centering
     \includegraphics[width=\textwidth]{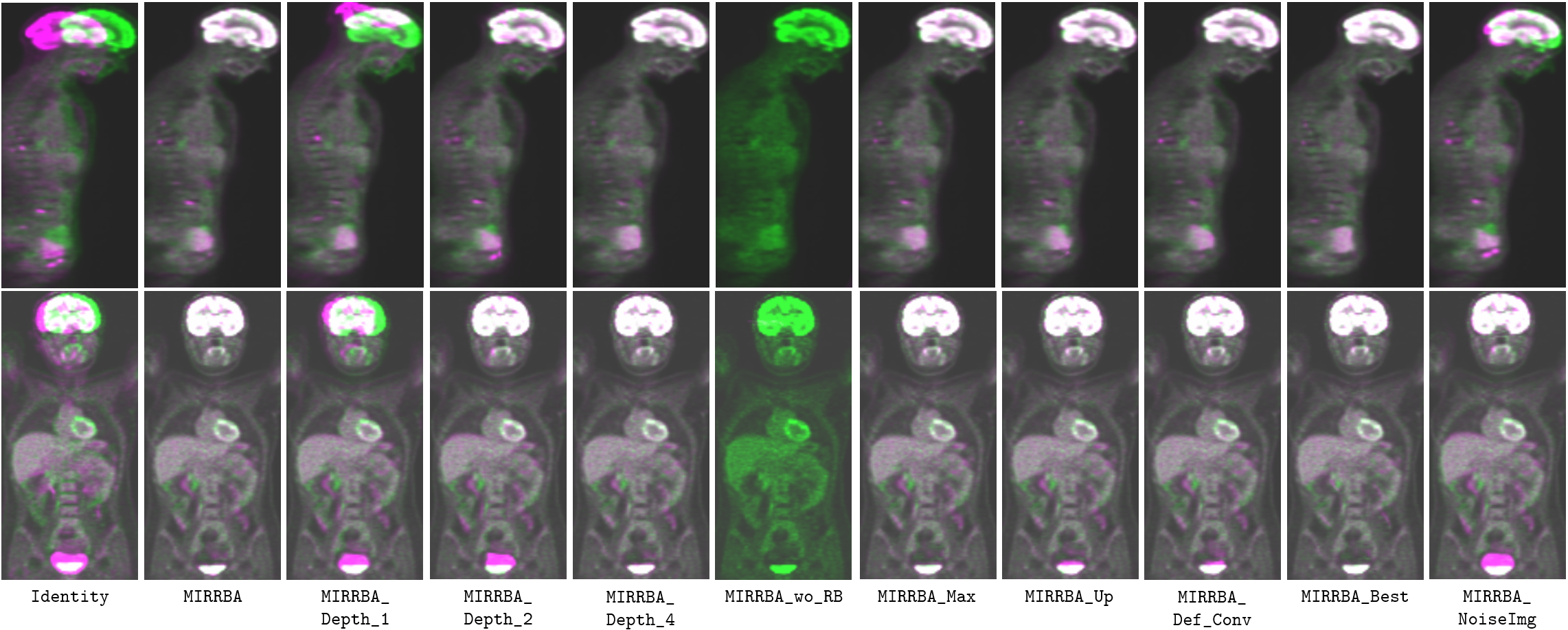}
     \caption{Ablation study -- Overlay of the fixed (green) and warped (pink) images on two different patients after performing \texttt{Identity} registration, \texttt{MIRRBA} (corresponding to depth 3), \texttt{MIRRBA\_Depth\_1}, \texttt{MIRRBA\_Depth\_2}, \texttt{MIRRBA\_Depth\_4}, \texttt{MIRRBA\_wo\_RB}, \texttt{MIRRBA\_Max}, \texttt{MIRRBA\_Up}, \texttt{MIRRBA\_DefConv}, \texttt{MIRRBA\_Best} and \texttt{MIRRBA\_NoiseImg}. Grayscale color indicates good overlapping. We can see that the higher the depth, the more precise the registration. A simple U-Net-shaped architecture as \texttt{MIRRBA\_Depth\_1} produces transformations of very low accuracy, while the more resolutions are used, the more precise the registration. The all green \texttt{MIRRBA\_wo\_RB} image is due to the non-convergence of the registration algorithm and a wrapped image not registered to the fixed one. All four other approaches produce realistic looking and coherent transformations, even if \texttt{MIRRBA\_DefConv} lacks a bit of precision around the bladder. \texttt{MIRRBA\_NoiseImg} does not reach local precise registration. Best viewed in color.}
     \label{fig:01N-008_01A-016_network}
\end{figure*}

\end{document}